\documentclass[3p,twocolumn]{elsarticle}

\usepackage{lineno}
\usepackage{amsmath}
\usepackage{cuted,multirow}
\modulolinenumbers[5]

\journal{Physics Letters A}

\begin{document}

\begin{frontmatter}

\title{Propagation of terahertz waves in a dust acoustic wave}


\author[mymainaddress]{Peng-Fei Li}

\author[mymainaddress,mysecondaryaddress]{Cheng-Ran Du\corref{mycorrespondingauthor}}
\cortext[mycorrespondingauthor]{Corresponding author}
\ead{chengran.du@dhu.edu.cn}

\address[mymainaddress]{College of Science, Donghua University, 201620 Shanghai, PR China}
\address[mysecondaryaddress]{Member of Magnetic Confinement Fusion Research Centre, Ministry of Education, 201620 Shanghai, PR China}

\begin{abstract}
The propagation of terahertz waves in a dust acoustic wave is investigated numerically. By assuming a sinus profile of the dust number density in the dust acoustic waves, the transmission properties are calculated using finite difference time domain method. It shows that the dust acoustic wave can function similarly as a Bragg filter to block the terahertz waves of a certain wavelength. The bandwidth of the filter depends on the density profile of the dust acoustic wave.
\end{abstract}

\begin{keyword}
dusty plasma \sep dust acoustic wave \sep terahertz wave
\end{keyword}

\end{frontmatter}

Dusty plasmas are composed of an ionized gas and dust particles, which acquire charges while interacting with ions and electrons \cite{Fortov:2005,Bouchoule:book}. They are widely found in planetary rings, interplanetary and interstellar clouds, cometary tails, the earth's mesosphere, thunderclouds, as well as in the vicinity of artificial satellite, spaceships, and space stations, etc \cite{Whipple:1981,Goertz:1989,Kempf:2005,Horanyi:2015}. Dusty plasmas are also present in the industrial apparatuses and fusion devices \cite{Selwyn:1989,Krasheninnikov:2011}. Recently, they are prepared in the laboratory on the ground \cite{Thomas:1996,I:1996} and on board the International Space Station \cite{Thomas:2008,Pustylnik:2016}, to study the structural and dynamical properties of solids and liquids as a model system \cite{Morfill:2009,Chaudhuri:2011,Ivlev:book}.

Dust acoustic wave (DAW) is one of the most important phenomena observed in the dusty plasmas \cite{Shukla:book,Rao:1990,Merlino:2014}. It is regarded as a possible mechanism for the fluctuations observed in many space and astrophysical systems \cite{Popel:2013,Ghosh:2001}. In the laboratory, DAW can be either self-excited, triggered by ion streaming instability \cite{Rosenberg:1996}, or be excited by external drivers, such as periodic electric fields \cite{Thompson:1997}. Since the dust particles are much larger and heavier than the electrons and ions, the frequency of DAW can be lowered to a few hertz. This enables a direct observation of the wave-particle interaction in the experiments using simple video microscopy \cite{Schwabe:2007,Yang:2017,Jaiswal:2018}.

\begin{figure}[!ht]
	\includegraphics[width=19pc]{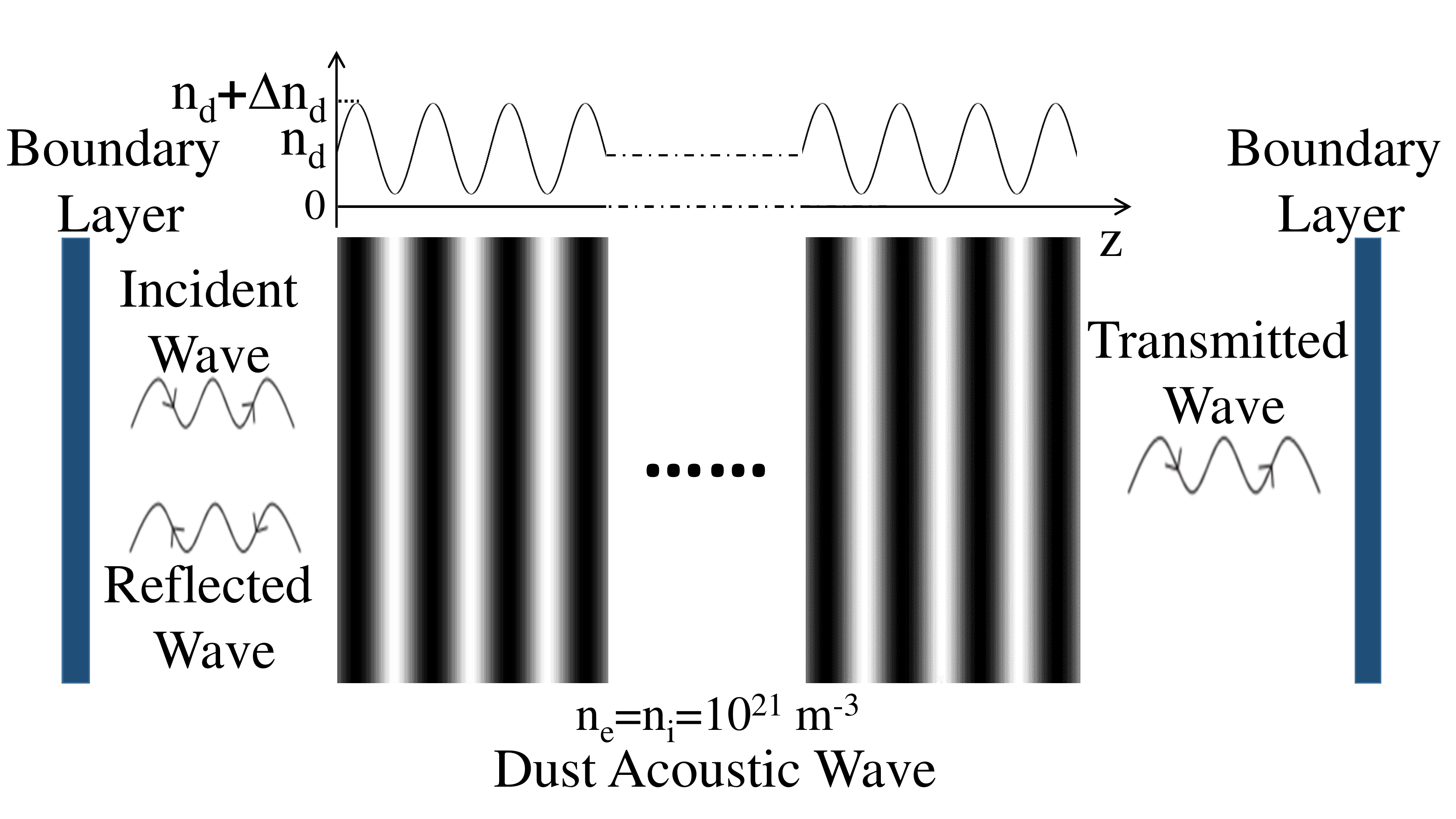}
	\caption{ Sketch of the simulation domain. The Gaussian pulse propagates from the left to the right. The ion and electron density are set as constant in the DAW. The periodic profile of the number density of the dust particles in the DAW is assumed to have a sinus shape.}
	\label{fig1}
\end{figure}

For the dusty plasma prepared in the laboratory, the size of the dust particles is typically about hundreds of nanometers to microns while the spacing can range from tens of microns to millimeters \cite{Du:2010}. This length scale is comparable to the wavelength of the terahertz (THz) electromagnetic (EM) waves, which have wide applications in science and industry \cite{Lee:book,Saeedkia:Book}. Because of this remarkable feature, lattice of dust particles in the plasma, known as plasma crystal \cite{Zuzic:2006}, can work as a filter for the THz waves \cite{Rosenberg:2006}. Furthermore, the propagation of THz waves in a homogeneous dusty plasma has been investigated theoretically \cite{Jia:2016,Wang:2015,Chen:2019,Jia:2015}. In this letter, we explore the propagation characteristics of THz waves in a dust acoustic wave using the finite-difference time-domain method \cite{Takahashi:2015}.

The dielectric permittivity of a dusty plasma can be derived based on the Boltzmann equation \cite{Jia:2015}. The dust particles immersed in the plasma are charged by absorbing electrons and ions. The charging process reaches an equilibrium as the ion and electron current to the surface of the particle balance off. Due to the higher thermal velocities of electrons, the overall charge of the particles are usually negative \cite{Fortov:2005}. The propagation of THz wave alters the charging current. Since the mass of ions is much higher than that of electrons, impact of THz wave on the ion current is neglected. In a fully ionized dusty plasma, the electron-ion collision is dominant. Considering the charging and collision effect, the real part of the dielectric permittivity $\varepsilon_d$ in the fully ionized dusty plasma \cite{Jia:2016} can be expressed as

\begin{strip}
\rule{\dimexpr(0.5\textwidth-0.5\columnsep-0.4pt)}{0.4pt}%
\rule{0.4pt}{6pt}
\par
\begin{equation}
 \varepsilon_d = \varepsilon_0 - \varepsilon_0\frac{\omega_p^2}{\omega^2+\nu_{ei}^2}
                 +\frac{\pi e^2 r_d^2 n_e n_d }{m_e} \frac{\omega}{k} \frac{(\nu_{ei}+\nu_{ch})}{(\omega^2+\nu_{ei}^2)(\omega^2+\nu_{ch}^2)}
                 \left(1+\frac{Z e^2}{6\pi \varepsilon_0 r_d k_B T_e}\right),
 \label{eq1}
\end{equation}
\par
\hfill\rule[-6pt]{0.4pt}{6.4pt}%
\rule{\dimexpr(0.5\textwidth-0.5\columnsep-1pt)}{0.4pt}
\end{strip}

\noindent where $\omega$ and $k$ are the frequency and wave number of the THz wave, $m_e$ is the mass of the electron, $k_B$ is the Boltzmann constant, and $\varepsilon_0$ is the vacuum permittivity. The rest of the symbols are explained in Tab.~\ref{tab1}. For simplicity, the absorption of the dusty plasma is not taken into account.

\begin{table}[h]
\centering
\begin{tabular}{ l l l}
\hline
parameter & abbrev. & value \\
\hline \hline
plasma frequency & $\omega_p$ [Hz] & $1.8 \times 10^{12}$\\
electron-ion & \multirow{2}*{$\nu_{ei}$  [Hz]} & \multirow{2}*{$8.4 \times 10^{11}$} \\
collision frequency & ~ & ~ \\
electron density & $n_e$ [m$^{-3}$] & $1.0 \times 10^{21}$ \\
ion density & $n_i$ [m$^{-3}$] & $1.0 \times 10^{21}$ \\
electron temperature & $T_e$ [K] & $5.8 \times 10^{4}$ \\
particle radius & $r_d$ [nm] & $200$ \\
charging frequency & $\nu_{ch}$ [Hz] & $2.0 \times 10^{9}$ \\
particle charge & $Z$ [$e$] & $1.5 \times 10^{3}$ \\
mean particle density & $n_d$ [m$^{-3}$] & $2.0 \times 10^{16}$ \\
\hline
\end{tabular}
\caption{Parameters of the dust acoustic wave.}\label{tab1}
\end{table}

The propagation of the THz waves in a DAW can be simulated using the finite difference time domain (FDTD) method. The typical frequency of the DAW is about a few hertz, much lower than that of the EM waves. To study the transient propagation property of the THz waves, one can assume the spatial distribution of the particle density in the DAW doesn't change, namely the DAW doesn't propagate in the time scale of the period of EM wave. The simulation domain enclosed by the boundary layers is depicted in Fig.~\ref{fig1}, where the left and right parts are the glass windows (made of SiO$_2$) and the middle part is the DAW. The glass window has a thickness of $2.5$~mm. The DAW is cylindrical with a length of $10$~mm and diameter of $3$~mm. The number density of the dust particle is assumed to have a sinus profile with the mean particle density $n_d$ and an amplitude $\Delta n_d$. The electron and ion density are set as constant in the DAW\footnote{In general, the ions and electrons follow the Boltzmann distributions in the wave potential. However, as the electron temperature is high, the relative magnitude is small and thus is neglected in this study.}. The incoming wave is set as a Gaussian pulse from the left end. The transmission is calculated on the right end. The simulation time is set as $300$~ps. The plasma parameters are listed in the Tab.~\ref{tab1}.

\begin{figure}[!ht]
	\includegraphics[width=18pc]{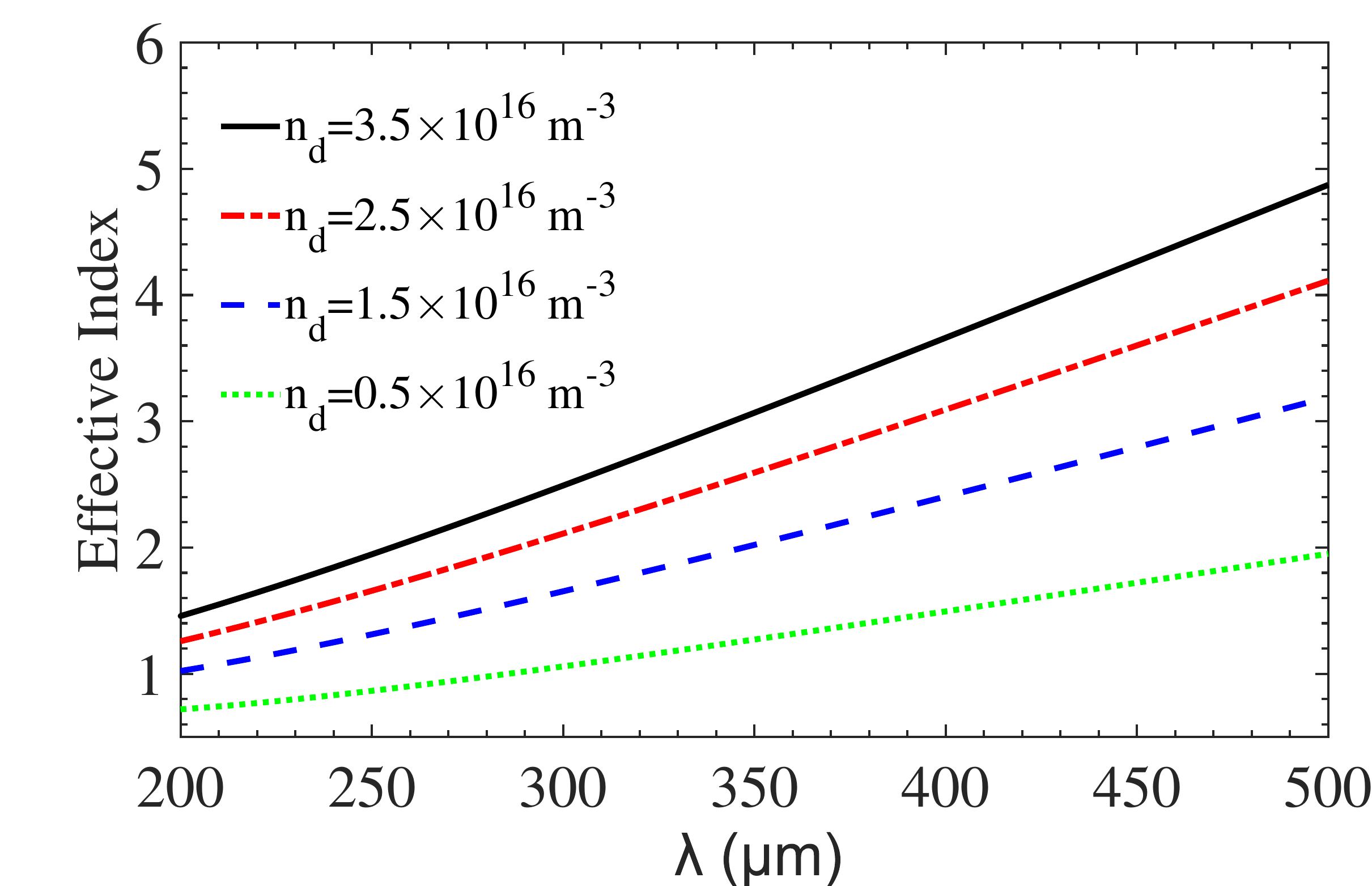}
	\caption{ Dependence of the effective refractive index on the wavelength of the THz waves and the number density of the dust particles.}
	\label{fig2}
\end{figure}

The dependence of the effective index on the number density of the dust particles is shown in Fig.~\ref{fig2}. The effective refractive index increases monotonically with the wavelength. Assuming a constant charge of the dust particles, the refractive index exhibits a linear relation to the number density of the dust particles. This results in a sinus profile of the refractive index in the DAW. Therefore, the DAW can work similarly as a Bragg filter that EM waves of certain wavelength are blocked.

\begin{figure}[!ht]
	\includegraphics[width=18pc]{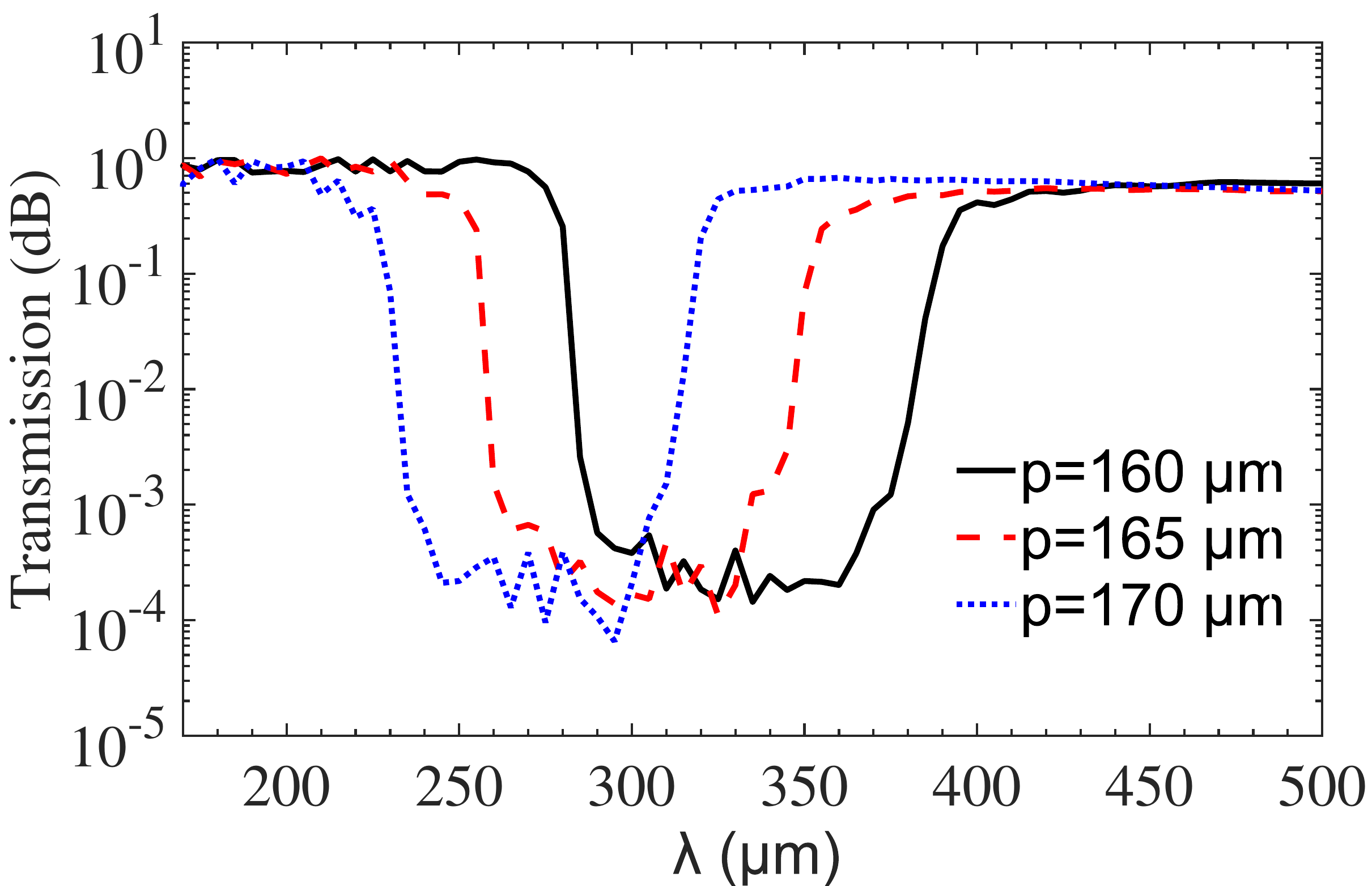}
	\caption{ Dependence of the transmission of the THz wave on the wavelength $p$ of the dust acoustic waves. The density amplitude of the DAW is  $\Delta n_d=0.7 n_d$.}
	\label{fig3}
\end{figure}

We study the propagation characteristics of the THz waves in the DAWs of three wavelengths $p=160$, $165$, and $170$~$\mu$m. The density amplitude $\Delta n_d$ in the DAW is $1.4\times10^{16}$~m$^{-3}$. The dependence of the transmission on the wavelength is shown in Fig.~\ref{fig3}. For $p=160$~$\mu$m, the transmission is close to unity for small wavelength and drops to $10^{-4}$ at $\lambda=280$~$\mu$m. The bandwidth is about $100$~$\mu$m. For the DAW of bigger wavelengths, the wavelength at which the transmission is blocked shifts to a smaller wavelength. For $p=165$ and $170$~$\mu$m, the central wavelength shifts from $340$~$\mu$m to $310$ and $280$~$\mu$m, respectively, with a comparable bandwidth.

\begin{figure}[!ht]
	\includegraphics[width=18pc]{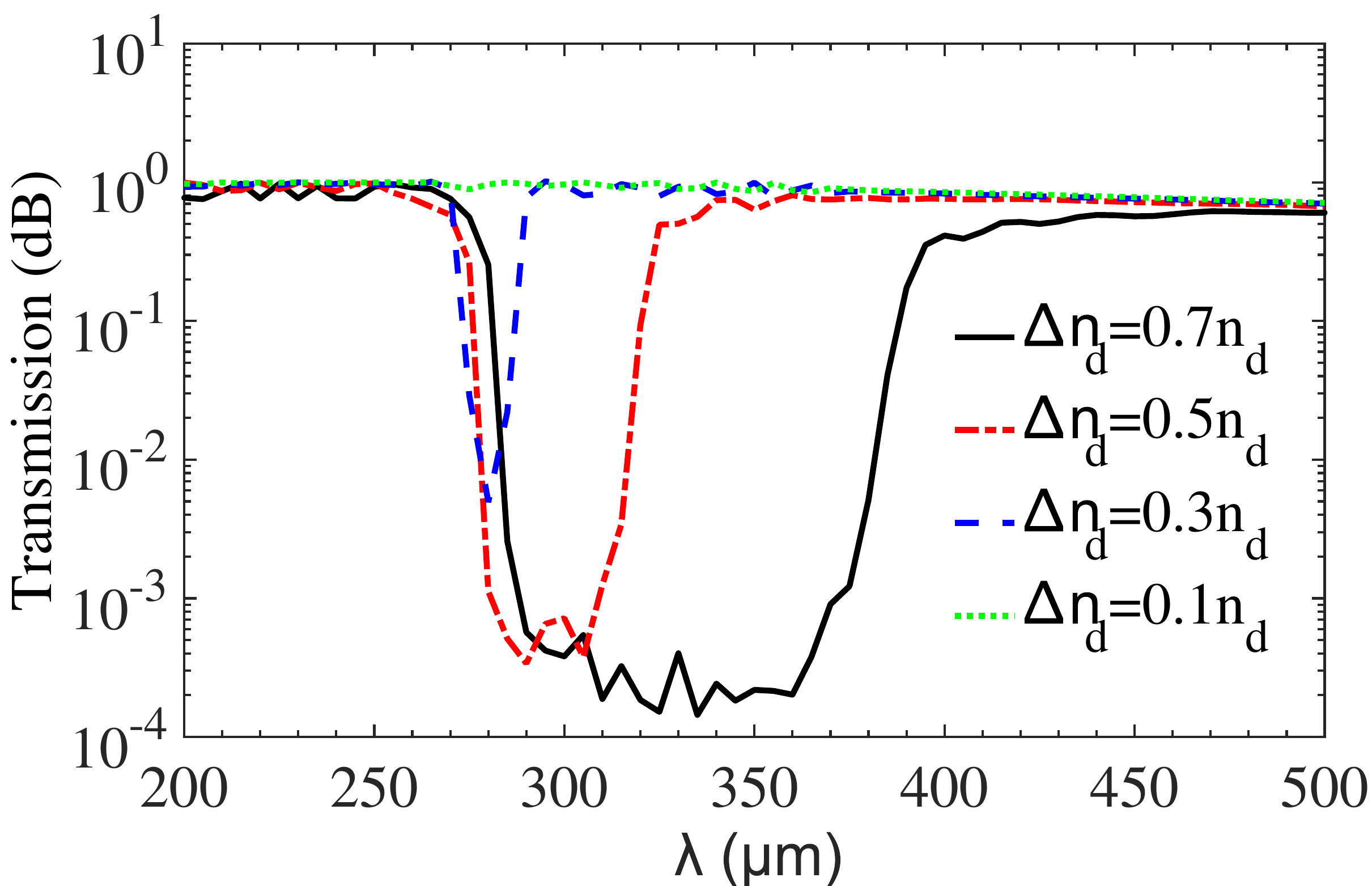}
	\caption{ Dependence of the transmission of the THz wave on the density amplitude $\Delta n_d$ of dust particles in the dust acoustic waves. The wavelength of the DAW is $p=160$~$\mu$m. }
	\label{fig4}
\end{figure}

We further study the dependence of the transmission of the THz wave on the density amplitude of the DAW. The wavelength of DAW is $160$~$\mu$m. As shown in Fig.~\ref{fig4}, the minimal transmission increases as the density amplitude decreases. For $\Delta n_d=0.3n_d$, the transmission decreases to $10^{-2}$ at $\lambda=280$~$\mu$m and the bandwidth shrinks to $10$~$\mu$m. As the $\Delta n_d$ further drops to $0.1n_d$, the transmission approaches unity for all the wavelength. The filtering effect can no longer be observed.

In conclusion, the transmission properties of THz waves in dust acoustic waves is investigated using FDTD method. For simplicity, we assume the sinus profile of the number density of the dust particles. The EM waves of certain wavelength are filtered depending on the wavelengths and density profile of the DAW. It turns out that the bandwidth depends on the density profile rather than the wavelengths. The results may shed light on the underlying mechanism of the signal blackout in the mesosphere, where the DAW is present.

However, this approach can only be applied to the cases where the dust particles are small in size. The scattering of the solid spherical particles in terms of their material and surface roughness are neglected. For dust particles of bigger sizes, these properties may play a significant role in the transmission characteristics of the THz waves in DAW \cite{Vladimirov:1994a,Vladimirov:1994b}. We leave this for future work.

The authors acknowledge the support from the National Natural Science Foundation of China (NSFC), Grant No.~$11975073$. We thank Sergey Khrapak for the valuable discussions.

\bibliography{reference}

\end{document}